\documentclass[aps,prd,twocolumn,superscriptaddress,nofootinbib,10pt]{revtex4-1}   

\usepackage[utf8]{inputenc}
\usepackage{amssymb}
\usepackage{amsmath}
\usepackage{amsfonts}
\usepackage{graphicx}
\usepackage{color}
\usepackage{xspace}
\usepackage{comment}
\usepackage{hyperref}
\usepackage[normalem]{ulem}

\usepackage[section]{placeins}
\usepackage{afterpage}

\usepackage{float}
\usepackage{slashed}
\usepackage{ulem}
\usepackage{appendix}

\usepackage{multirow,rotating}
\usepackage[dvipsnames]{xcolor}

\begin{document}
	
	%\preprint{}
        \title{Constraining long-lived particles from Higgs boson decays \\ at the LHC with displaced vertices and jets}

	\author{Zeren Simon Wang}
	\email{wzs@mx.nthu.edu.tw}
	\affiliation{Department of Physics, National Tsing Hua University, Hsinchu 300, Taiwan}
	\affiliation{Center for Theory and Computation, National Tsing Hua University, Hsinchu 300, Taiwan}

	%%%%%%%%%%%%%%%%%%%%%%%%%%%%%%%%%%%%%%%%%%%%%%%%%%%%%%%%%%%%%%%%%%%%%%
	\begin{abstract}       
           
Long-lived particles (LLPs) originating from decays of Standard-Model-like or Beyond-the-Standard-Model Higgs bosons are often featured with signatures of displaced vertices (DVs) and jets at colliders.
In this work, we show that a recent ATLAS search for DVs plus jets, with its recast implementation, can efficiently place bounds on such hadronically or semi-leptonically decaying LLPs.
In particular, we find the search is uniquely sensitive to LLP proper decay lengths of about 1-100 mm, probing complementary regions in the parameter space of the relevant models compared to other prompt and LLP searches.

 	\end{abstract}


	\vskip10mm
	
	\maketitle
	\flushbottom

\section{introduction}\label{sec:intro}

In recent years, searches for long-lived particles (LLPs) have become an increasingly important field in the quest for physics beyond the Standard Model (BSM).
This is due to not only the absence of discovery of any BSM fundamental heavy fields at the LHC, but also the fact that LLPs are widely predicted in various BSM models and often well motivated for different reasons including explaining non-vanishing active-neutrino masses or dark matter (DM).
Among the many possible theories predicting LLPs, a class of ``portal-physics'' models are particularly appealing for their simple constructions, strong predictability, as well as close connection to the DM.
These models propose a new field as the mediator connecting the visible sector (which is essentially the spectrum of the Standard Model (SM)) and a so-called ``hidden sector'' consisting of unknown, dark particles.
Typical examples of such a mediator include a dark scalar boson~\cite{OConnell:2006rsp,Wells:2008xg,Bird:2004ts,Pospelov:2007mp,Krnjaic:2015mbs,Boiarska:2019jym}, a dark photon~\cite{Okun:1982xi,Galison:1983pa,Holdom:1985ag,Boehm:2003hm,Pospelov:2008zw}, a heavy neutral lepton (HNL)~\cite{Shrock:1980vy,Shrock:1980ct,Shrock:1981wq} (see also Ref.~\cite{Abdullahi:2022jlv} for a recent review), and an axion-like particle~\cite{Peccei:1977hh,Peccei:1977ur,Witten:1984dg,Conlon:2006tq,Arkani-Hamed:2006emk,Arvanitaki:2009fg,Cicoli:2012sz}.
See Refs.~\cite{Alimena:2019zri,Lee:2018pag,Curtin:2018mvb,Beacham:2019nyx} for recent reviews on theories and experimental searches for LLPs.
In different BSM theories, various production mechanisms exist for these mediator particles, and in this work, we focus on their production from decays of the SM-like (or a BSM) Higgs boson.
Specifically, we choose to restrict ourselves to the dark scalar bosons and the HNLs that could be thus produced at the LHC.

In various BSM scenarios such as the SM extended by a singlet scalar that mixes with the SM-like Higgs boson~\cite{Silveira:1985rk}, neutral-naturalness models~\cite{Chacko:2005pe,Burdman:2006tz,Craig:2015pha,Cai:2008au,Cohen:2018mgv,Cheng:2018gvu} that solve the little hierarchy problem, and two-Higgs-doublet models~\cite{Lee:1973iz} (see Ref.~\cite{Branco:2011iw} for a review) including the supersymmetric models~\cite{Nilles:1983ge,Haber:1984rc,Martin:1997ns}, additional scalar particles are predicted.
Given these models' various kinds of strong motivation and the currently relatively loose bounds on the Higgs-boson properties, it is of much interest to study the SM-like (or BSM) Higgs-boson decays into a pair of light scalars.
In this work, we will treat the decay branching ratio of this channel and the proper decay length of the light scalar, as independent parameters, and present numerical results in planes spanned by these and similar variables, for various mass values of the hypothetical light scalar particle.
On the other hand, in $U(1)$ extensions of the SM such as $U(1)_{B-L}$~\cite{Davidson:1978pm,Mohapatra:1980qe} that explain the light neutrino mass, additional particles including the HNLs are predicted.
In particular, the mixing of the SM-like Higgs boson with the new scalar particle allows the Higgs boson to decay into a pair of HNLs that may further decay via (tiny) mixing parameters with the active neutrinos.
Here, the production and decay of the HNLs are decoupled, and we will fix the production rates according to existing bounds and show numerical results in the plane of the HNL mass and mixing angle.
In both scenarios, if the mass of the light scalar boson or the HNL is small, or their couplings to the SM particles are tiny, they naturally become long-lived, possibly circumventing traditional collider searches.

At high-energy proton-proton colliders such as the LHC, the Higgs bosons are dominantly produced in the gluon-fusion channel, and decay promptly.
If they decay to the long-lived light scalars or HNLs, these LLPs' displaced or delayed decays could give rise to signatures of displaced jets and displaced vertices at the LHC.
Recently, an ATLAS search~\cite{ATLAS:2023oti} with the full Run-2 dataset of 139 fb$^{-1}$ at the center-of-mass energy $\sqrt{s}=13$ TeV has been recast in Ref.~\cite{Cheung:2024qve}.
The search targets long-lived electroweakinos predicted in the R-parity-violating supersymmetry, featured with signatures of displaced vertices and jets (``DV plus jets'').
It starts with event- and vertex-level acceptance requirements imposed mainly on the jets and displaced vertices.
Besides, the recast relies on the usage of parameterized efficiencies also at both event- and vertex-levels provided by the ATLAS collaboration~\cite{ATLAShepdata}, for taking into account more complicated experimental selections such as multi-jet trigger and material effects.
While no discovery was made, new bounds on the masses and lifetimes of the long-lived electroweakinos were established.
Given the similar signatures, we choose to reinterpret the obtained bounds in terms of the long-lived light scalar or HNL originating from decays of the SM-like or a BSM Higgs boson produced in gluon-fusion processes\footnote{We have explicitly checked the case of vector-boson-fusion (VBF) Higgs-boson production. While VBF processes have extra prompt jets, the cutflow efficiencies are similar to those observed with gluon-fusion processes. However, the former processes' cross sections are smaller by more than one order of magnitude, and as a result, the final sensitivities in all the theoretical scenarios considered in this work would be inferior to different degrees. Therefore, we choose to restrict ourselves to the gluon-fusion production modes of the Higgs bosons.}.
We will show that this search can efficiently constrain these LLP scenarios, particularly in the $\mathcal{O}(1\text{ - }10)$ mm range of the proper decay length of the LLPs, probing unique regions of the parameter space that are hard to access with other searches.

We organize the paper as follows.
In Sec.~\ref{sec:search}, we discuss the ATLAS DV+jets search~\cite{ATLAS:2023oti} and review its recast implementation first presented in Ref.~\cite{Cheung:2024qve}.
Then in Sec.~\ref{sec:models} we introduce the theoretical scenarios we focus on, including the corresponding relevant signal processes.
We proceed to present the numerical results in Sec.~\ref{sec:results} for the considered theoretical scenarios.
Finally, we summarize our findings and conclude the work in Sec.~\ref{sec:conclusions}.
Additionally, we show in Appendix~\ref{appendix:bbbar} a plot for the decay branching ratio of a Higgs-like light scalar $\phi$ into a pair of $b$-quarks as a function of its mass $m_\phi$, which is required for the numerical estimates in the first theoretical scenario we will investigate.

\section{The ATLAS DV+jets search and the recast}\label{sec:search}

The ATLAS collaboration reported a search for massive, multi-track DVs and multiple jets~\cite{ATLAS:2023oti}.
Making use of a dataset of an integrated luminosity of 139 fb$^{-1}$ collected at the center-of-mass energy $\sqrt{s}=13$ TeV, the search aims at long-lived electroweakinos ($\tilde{\chi}^0_1$, $\tilde{\chi}^0_2$, and $\tilde{\chi}_1^\pm$) in the R-parity-violating supersymmetry which decay via $\lambda'' \bar U \bar D \bar D$ operators into jets, in alignment with the considered signatures.
No discovery was made in the search, but exclusion bounds on the masses and lifetimes of the long-lived electroweakinos were established.

In the following, we provide a description of the recasting procedure.
The search defines two signal regions (SRs), viz.~High-$p_T$-jet SR and Trackless-jet SR, and the event selections in both SRs start with requirements on the transverse momentum $p_T$ of truth jets, as detailed in Table~\ref{tab:event_acceptance}.
\begin{table}[t]
\begin{center}
\resizebox{0.48\textwidth}{!}{%
\begin{tabular}{c|c|c}
SR            & High-$p_T$ jet                                                  & Trackless jet                                                                  \\ \hline
              & $n^{250}_{\text{jet}}\geq 4$ or $n^{195}_{\text{jet}}\geq 5$    & $n^{137}_{\text{jet}}\geq 4$ or $n^{101}_{\text{jet}}\geq 5$                   \\
Jet selection & or $n^{116}_{\text{jet}}\geq 6$  or $n^{90}_{\text{jet}}\geq 7$ & or $n^{83}_{\text{jet}}\geq 6$ or $n^{55}_{\text{jet}}\geq 7$,                    \\
              &                                                                 & $n^{70}_{\text{disp.~jet}}\geq 1$ or $n^{50}_{\text{disp.~jet}}\geq 2$ \\ \hline
\end{tabular}
}
\caption{Selection requirements on the truth jets. Here, $n^{250}_{\text{jet}}$ denotes the number of jets with a $p_T$ larger than or equal to 250 GeV, and the other notations employ similar meanings.
Plus, ``disp.~jet'' stands for ``displaced jet''.}
\label{tab:event_acceptance}
\end{center}
\end{table}
Here, the ``displaced jets'' are the jets that are determined to have originated from the decay of an LLP by checking $\Delta R$ between the LLP's decay products and the truth jet.
The proportion of events having passed these event selections defines the event-level acceptance.
The search then proceeds to vertex-level selections which require that the events should include at least one vertex that fulfills the conditions listed below,
\begin{enumerate}
    \item $R_{xy}, |z| < 300$ mm where $R_{xy}$ and $|z|$ are the absolute distance of the vertex to the IP in the radial and longitudinal directions, respectively.
    \item $R_{xy}>4$ mm.
    \item At least one track stemming from the LLP decays should have an absolute transverse impact parameter $|d_0|$ larger than 2 mm.
    \item At least 5 massive decay products from the DV should exist that pass the following two requirements:  
    \begin{enumerate}
        \item The decay product should be a track, with a lab-frame decay length in the radial direction ($\beta_t \gamma c\tau$) larger than 520 mm, where $\beta_t$ is the absolute speed of the decay product in the radial direction, $\gamma$ is the boost factor, and $c\tau$ is the proper decay length.
        \item The decay product should have $p_T$ and electric charge $q$ satisfying $p_T/|q|> 1$ GeV.
    \end{enumerate}
    \item $m_{\text{DV}}> 10$ GeV with $m_{\text{DV}}$ labeling the invariant mass of the DV. Note that $m_{\text{DV}}$ is computed with the decay products satisfying the above conditions, and the mass of these decay products is assumed to be the same as that of a charged pion.
\end{enumerate} 
Following the event- and vertex-level acceptances discussed above, the recast of the search relies on application of sets of parameterized efficiencies relevant at both event- and vertex-levels provided by the ATLAS collaboration on the HEPData website~\cite{ATLAShepdata}.
Event-level efficiencies depend on the LLP decay position and the sum of the truth jets' $p_T$, and vertex-level efficiencies are functions of the LLP decay positions, $m_{\text{DV}}$, as well as the number of tracks associated with a truth decay vertex.
These efficiencies are purposed for accounting for further and complicated search selections that are difficult to simulate, such as multi-jet trigger, High-$p_T$/Trackless-jet filter, and material effects.

Background events mainly originate from erroneous merge of nearby DVs of small invariant masses by vertexing algorithms resulting in a high-mass DV, hadronic interactions between particles and detector materials, as well as accidental crossings of a track with unrelated low-mass DVs.
After all the above-mentioned event selections are applied, $\mathcal{O}(1)$ background events are expected.
Systematic uncertainties on the expected background-event numbers arising from various sources including pile-up effects are also estimated.
Finally, the expected background-event numbers are thus assessed to be $0.46^{+0.27}_{-0.30}$ and $0.83^{+0.51}_{-0.53}$ at the High-$p_T$-jet and Trackless-jet SRs, respectively.
The search observed 1 and 0 events in these SRs at the end.
Given the almost vanishing expected background contamination and numbers of observed events after all the event selections, the search derives new-physics signal-event numbers of 3.8 and 3.0 corresponding to the exclusion bounds at 95\% confidence level (C.L.) for the High-$p_T$-jet and the Trackless-jet SRs, respectively, assuming little contamination from the new physics on the background-level predictions.
Further, we note that when we present numerical results later, despite the expected higher levels of pile-up events in the HL-LHC (3000 fb$^{-1}$ integrated luminosity), we optimistically assume that future advancements in technologies, experimental search algorithms, etc., will allow to achieve the same level of background events with the same analysis, and we will, therefore, take the same numbers of the signal-event numbers as the sensitivity reach at 95\% C.L.

For more detail of the search and the recasting procedure, see Refs.~\cite{ATLAS:2023oti,ATLAShepdata,Cheung:2024qve}.

\section{Theoretical models and signal process}\label{sec:models}

In this section, we introduce the theoretical models, their associated LLPs, as well as the signal processes.

\subsection{A light singlet scalar from SM-like Higgs boson decays}\label{subsec:lightscalar}

We start with a model where an extra scalar particle, $\phi$, mixes with the SM-like Higgs boson.
It can be pair-produced from rare decays of the SM-like Higgs boson, and decay to a pair of leptons, jets, or gauge bosons, depending on its mass.
Here, we are interested in the mass range roughly between 10 GeV and 62 GeV, where the lower reach of 10 GeV is determined by the invariant-mass cut applied in the ATLAS DV+jets search and the upper reach of 62 GeV corresponds approximately to the kinematic threshold of $m_h/2$ with $m_h$ denoting the mass of the SM-like Higgs boson.
In this mass range, the dominant decay mode of the light scalar is $\phi \to b\bar{b}$ and we therefore take it as the signature final state:
\begin{eqnarray}
pp\xrightarrow{\text{g.f.}} h \to \phi \phi,  (\phi\to b \bar{b}, \phi\to b \bar{b}),
\end{eqnarray}
where ``g.f.'' stands for ``gluon fusion''.
For the decay branching ratio of the light scalar into a pair of $b$-quarks, Br$(\phi\to b\bar{b})$, we assume the new scalar particle is a SM-like scalar particle with a mass different from that of the SM-like Higgs boson discovered at the LHC, and we use the program HDECAY 3.4~\cite{Djouadi:1997yw,Djouadi:2018xqq} to compute it for the considered mass range of the light scalar.

We work with model-independent observables here: Br$(h\to \phi\phi)$, $c\tau_\phi$, and $m_\phi$, where $c\tau_\phi$ and $m_\phi$ are the proper decay length and the mass of $\phi$, respectively.
We note that the cross section of the Higgs-boson production in the gluon-fusion process at $\sqrt{s}=13$ TeV is $\sigma_h^{\text{g.f.}}\approx 48.5$ pb~\cite{ggf_higgs_xs}.

\subsection{A heavy Higgs boson and long-lived light scalar}\label{subsec:heavyHiggs}

In this case, we study a BSM heavy Higgs boson $\Phi$ also produced in gluon fusion at the LHC.
It then decays to a pair of light scalar bosons $s$.
This scenario has been targeted in various experimental searches involving a long-lived $s$ (see e.g.~Refs.~\cite{ATLAS:2019jcm,ATLAS:2019qrr,ATLAS:2022zhj}).
We thus restrict us to certain benchmark points reproduced from these experimental studies.
Concretely, we follow Refs.~\cite{ATLAS:2019jcm,ATLAS:2019qrr,ATLAS:2022zhj} to assume that the light scalar $s$ decays into $b\bar{b}$, $c\bar{c}$, and $\tau \bar{\tau}$ final states, with a fixed branching ratio of 85\%, 5\%, and 8\%, respectively, and its decay widths into bosons or a top-quark pair are vanishing.
We focus on the following three combinations of the masses of $\Phi$ and $s$ for study: $(m_\Phi, m_s)=  (600\text{ GeV}, 150\text{ GeV})$, $(1000\text{ GeV}, 275\text{ GeV})$, and $(1000\text{ GeV}, 400\text{ GeV})$.
We will present the numerical results in the plane spanned by $c\tau_s$ and $\sigma(\Phi)\cdot\text{Br}(\Phi\to ss)$, for each scalar-mass combination, where $c\tau_s$ is the proper decay length of $s$ and $\sigma(\Phi)$ is the production cross section of the $\Phi$ particle through gluon fusion.

\subsection{Heavy neutral leptons from SM-like Higgs boson decays}\label{subsec:HNL_SMHiggs}

We consider a UV-complete model, $U(1)_{B-L}$~\cite{Davidson:1978pm,Mohapatra:1980qe}, where the SM gauge group is extended by a $U(1)_{B-L}$ group.
We follow Refs.~\cite{Deppisch:2019kvs,Accomando:2016rpc,Deppisch:2018eth,Das:2018tbd,Das:2019fee,Amrith:2018yfb,Chiang:2019ajm,Liu:2022kid,Liu:2023klu,Liu:2024fey} for the particular model variant we will be using for the collider analysis here.
The model predicts three HNLs $N$, a new heavy gauge boson $Z'$, and a new Higgs boson $H$.
The new Higgs particle is required in order to break the $U(1)_{B-L}$ symmetry to generate a Majorana mass for the HNLs, and shall mix with the SM-like Higgs boson $h$, with a small mixing angle $\alpha$.
Light active-neutrino masses are then generated via seesaw mechanisms~\cite{Minkowski:1977sc,Yanagida:1979as,Gell-Mann:1979vob,Mohapatra:1979ia,Schechter:1980gr,Wyler:1982dd,Mohapatra:1986bd,Bernabeu:1987gr,Akhmedov:1995ip,Akhmedov:1995vm,Malinsky:2005bi}.
In this phenomenological analysis, we will treat the HNL masses and the active-sterile neutrino mixing angles as independent parameters.

The HNLs can be produced from the SM-like Higgs boson's decays via the small mixing $\alpha$, with $h\to N N$~\cite{Accomando:2016rpc,Deppisch:2018eth}.
The corresponding decay width for one HNL generation can be computed with~\cite{Accomando:2016rpc}
\begin{eqnarray}
	\Gamma(h\rightarrow NN) = \frac{1}{2}\frac{m_N^2}{\tilde{x}^2}\sin^2{\alpha}\,\frac{m_h}{8\pi}\,\Big( 1- \frac{4 m_N^2}{m_h^2} \Big)^{3/2},\label{eqn:Gammah2NN}
\end{eqnarray}
where $\tilde{x}=m_{Z'}/2g'_1$ is the vacuum expectation value (vev) of the new scalar boson with $m_{Z'}$ being the mass of the predicted $Z'$-boson and $g'_1$ being the gauge coupling of the appended $U(1)_{B-L}$ gauge group.
We emphasize that we assume only one generation of the HNLs are kinematically relevant while the other two are so heavy as to be decoupled.
For simplicity, we also assume that the HNL $N$ only mixes with the electron active neutrino.
Eq.~\eqref{eqn:Gammah2NN} allows to compute the Higgs-boson decay branching ratio into a pair of the HNLs with,
\begin{eqnarray}
	\text{Br}(h\rightarrow NN) = \frac{\Gamma(h\rightarrow NN)}{\Gamma(h\rightarrow NN) + \cos^2{\alpha}\,\Gamma^h_{\text{SM}}},\label{eqn:Brh2NN}
\end{eqnarray}
where $\Gamma^h_{\text{SM}}=4.1$ MeV~\cite{ParticleDataGroup:2022pth} labels the total decay width of the SM Higgs boson.

In particular, we take the following benchmark for the model parameters~\cite{Deppisch:2018eth,Chiang:2019ajm}:
\begin{eqnarray}
	m_N &=& 12\text{ - }62 \text{ GeV}, |U_{e N}|^2=10^{-12}\text{ - }10^{-6},\nonumber\\
	m_{Z'}&=&6 \text{ TeV}, g'_1=0.8, \tilde{x}=3.75 \text{ TeV},
	\label{eqn:U1BmL_parameters}\\
	&&m_{H}=450 \text{ GeV}, \sin{\alpha}=0.3, \nonumber
\end{eqnarray}
where $m_H$ is the mass of the new scalar particle.
The allowed benchmark values of $m_{Z'}$ and $g'_1$ were proposed in Ref.~\cite{Chiang:2019ajm} and are still allowed, and the bounds on $m_H$ and $\sin{\alpha}$ can be found in Ref.~\cite{Amrith:2018yfb}.

To compute the production cross section of the HNLs, we make use of the following expression:
\begin{eqnarray}
	\sigma(pp\xrightarrow{\text{g.f.}} h \rightarrow N N) = \cos^2{\alpha}\cdot  \sigma_h^{\text{g.f.}}\cdot \text{Br}(h\rightarrow NN). \label{eqn:pp2h2NN_XS}
\end{eqnarray}

Finally, we follow Ref.~\cite{DeVries:2020jbs} to compute the decay widths of the Majorana HNLs.
In the numerical analysis, we will focus on HNL decays into a neutrino or a charged lepton, plus two jets (including the $u$, $d$, $c$, $s$, and $b$-quarks, and their anti-particles), potentially resulting in the DV+jets signature.
We note that in this model, the production of the HNLs are mediated by the scalar-mixing and the vev of the new scalar boson, while their decay is induced separately by the active-sterile-neutrino mixing parameter $|U_{eN}|^2$.

\section{Numerical results}\label{sec:results}

In order to obtain the numerical results, we make use of the Monte-Carlo (MC) simulation tool Pythia 8.308~\cite{Bierlich:2022pfr}, and scan over the mass and lifetime of the LLPs in the considered scenarios by running one million signal events at each grid point.
Specifically, we utilize the module \textit{HiggsSM:gg2H} for simulating the gluon-fusion production of either the SM-like Higgs boson or the heavy BSM scalar (for the latter we tune the mass of the simulated SM Higgs boson) at the LHC with $\sqrt{s}=13$ TeV.
The generated Higgs bosons are forced to decay into a pair of new scalars or fermions with 100\% BR.
The mass and lifetime of the new scalar or fermion are correspondingly set up, and the simulated LLPs also exclusively decay to the signature final states only with the relative branching ratios of the decay channels correctly implemented.
These setups allow for achieving the largest possible amount of statistics, in order to properly evaluate the analysis efficiencies in an optimal way.
Finally, taking into account the production cross sections of these LLPs ``$\sigma(\text{LLP})$'', the integrated luminosities ($\mathcal{L}=139$ fb$^{-1}$ or 3000 fb$^{-1}$), as well as the signature branching ratios of the LLPs ``$\text{Br}(\text{sig.})$'', we compute the signal-event numbers in each case with the following formula:
\begin{eqnarray}
    N_S  = \sigma(\text{LLP})\cdot \mathcal{L}\cdot \epsilon \cdot \Big(\text{Br}(\text{sig.})\Big)^2,
\end{eqnarray}
where $\epsilon$ denotes the final cutflow efficiencies of the ATLAS DV+jets search, and Br$(\text{sig.})=\text{Br}(\phi\to b \bar{b}), \text{Br}(s\to b\bar{b}, c\bar{c}, \tau\bar{\tau}),\text{ and }\text{Br}(N\to e/\nu_e \, j \, j )$, for the three theoretical scenarios, respectively.
The power of 2 on Br$(\text{sig.})$ is due to the fact that in each signal event \textit{both} LLPs should decay to the specified signature final states.

Finally, we mention that we will follow the approach taken in Ref.~\cite{Cheung:2024qve} that the predicted signal-event numbers have an uncertainty of 50\% with the implemented recast analysis, and the sensitivity results are all shown with corresponding error bands.

\subsection{Light scalars from the SM-like Higgs boson}\label{subsec:results_lightscalar}

\begin{figure}[t]
    \centering
    \includegraphics[width=0.495\textwidth]{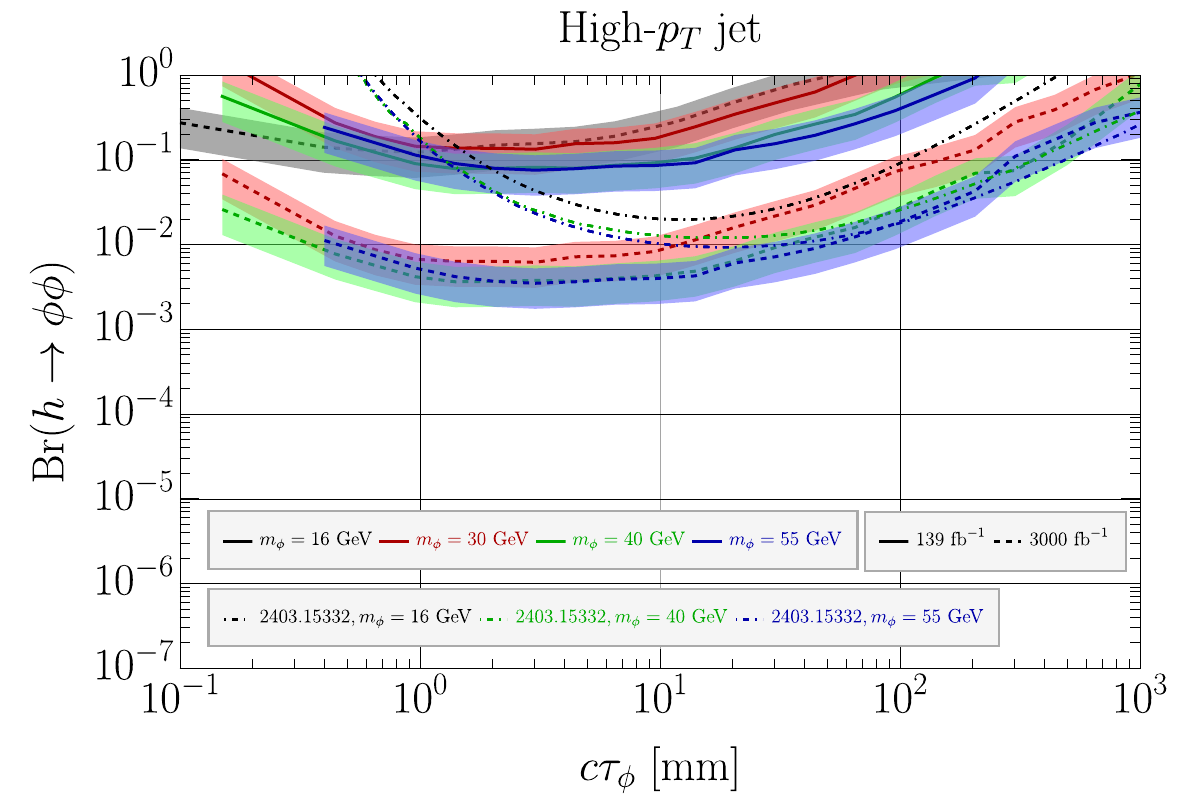}
    \includegraphics[width=0.495\textwidth]{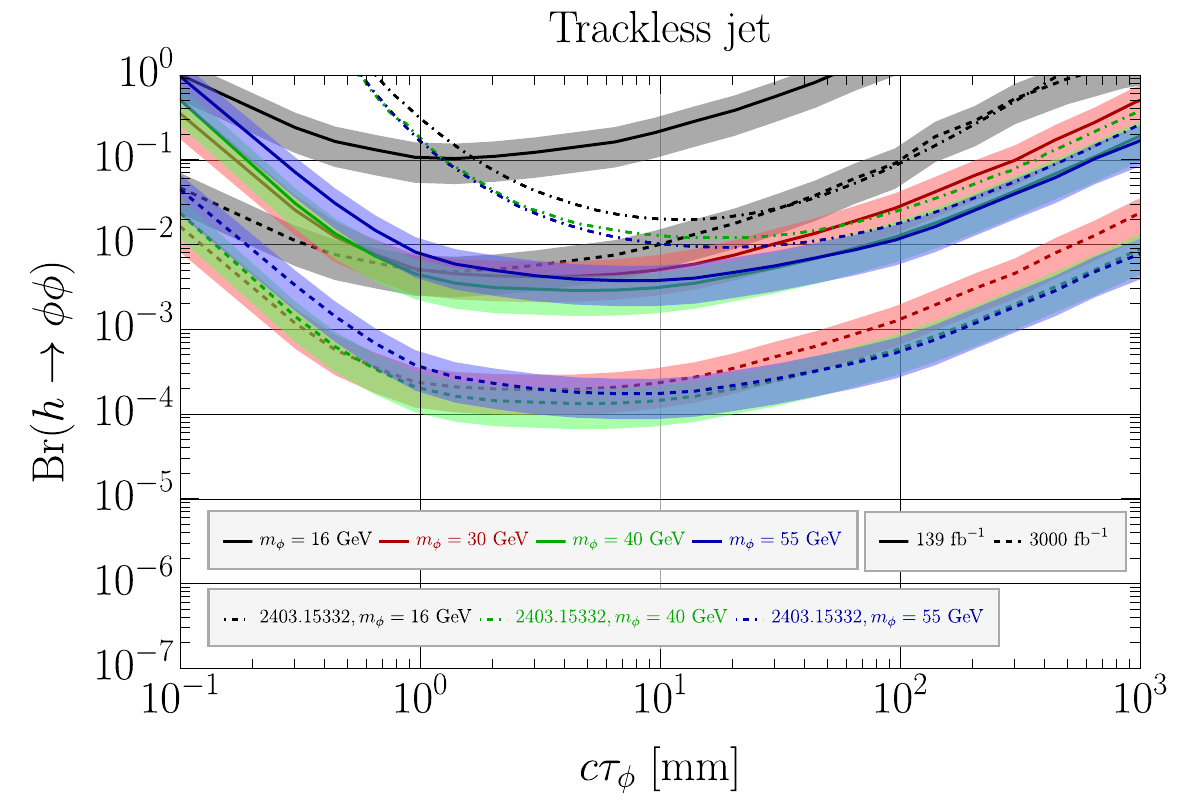}
    \caption{Sensitivity reach of the two SRs in the ATLAS DV+jets search to Br$(h\to \phi\phi)$ vs.~$c\tau_\phi$ for $m_{\phi}=16$, 30, 40, and 55 GeV.
    The error bands correspond to an uncertainty of 50\%; the same convention in the remaining sensitivity plots of this paper is taken.
    }
    \label{fig:Br_ctau_lightscalar}
\end{figure}

\begin{figure}[t]
    \centering
    \includegraphics[width=0.495\textwidth]{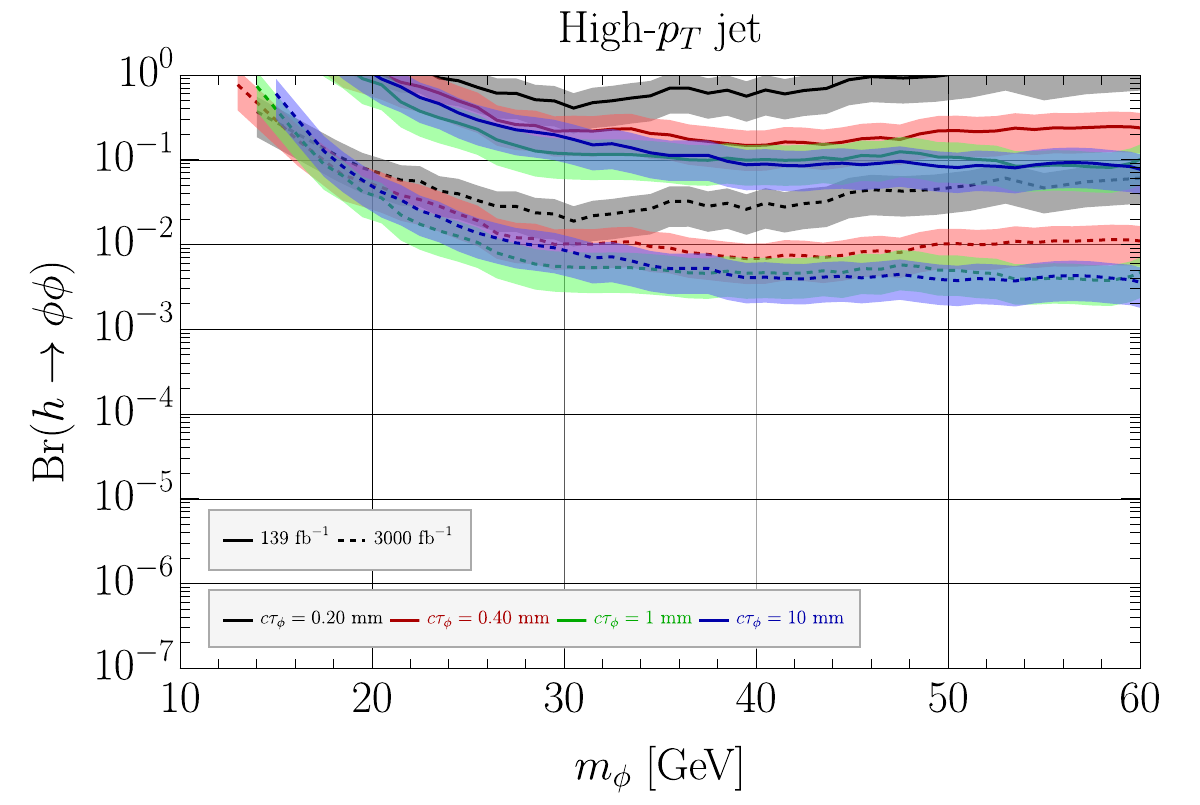}
    \includegraphics[width=0.495\textwidth]{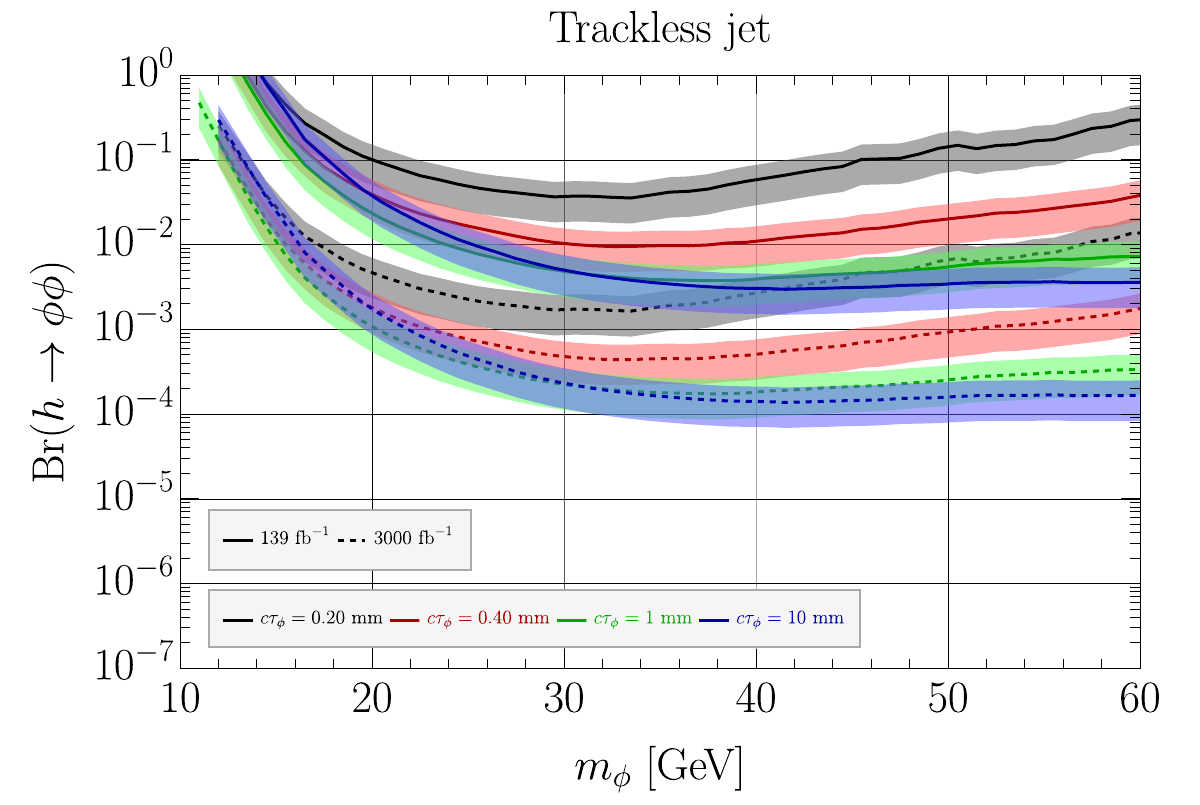}
    \caption{Sensitivity reach of the two SRs in the ATLAS DV+jets search to Br$(h\to \phi\phi)$ vs.~$m_{\phi}$ for $c\tau_\phi=0.20$, 0.40, 1, and 10 mm.}
    \label{fig:Br_mass_lightscalar}
\end{figure}

We present the numerical results for this scenario in the plane Br$(h\to \phi \phi)$ vs.~$c\tau_\phi$ and Br$(h\to\phi\phi)$ vs.~$m_\phi$ in Fig.~\ref{fig:Br_ctau_lightscalar} and Fig.~\ref{fig:Br_mass_lightscalar}, respectively.
Both figures contain two plots, corresponding to the results in the High-$p_T$-jet and the Trackless-jet SRs.
We display not only the recast current bounds from the published ATLAS DV+jets search with 139 fb$^{-1}$ integrated luminosity (in solid line style) but also the projected sensitivity reach of the same search with 3000 fb$^{-1}$ data volume (in dashed line style).

In Fig.~\ref{fig:Br_ctau_lightscalar}, we choose benchmark values of $m_\phi=16, 30, 40,$ and 55 GeV.
We observe, firstly, that the Trackless-jet SR is stronger in probing Br$(h\to \phi\phi)$ by roughly an order of magnitude than the High-$p_T$-jet SR.
This is mainly because in this scenario the light scalars are produced from the SM-like Higgs boson of which the mass is not so large compared to the jet-$p_T$ requirements in the event-level acceptance requirements of the ATLAS DV+jets search, cf.~Table~\ref{tab:event_acceptance}, especially so in the High-$p_T$-jet SR.
We also find that the bounds for the $m_\phi=16$ GeV case are much weaker than those for the heavier-$\phi$ cases, and in particular, in the High-$p_T$-jet SR the current bounds on Br$(h\to\phi \phi)$ for this mass choice are above 100\% and are hence not shown.
This arises because of the selection requirement in the ATLAS search on the DV invariant mass: $m_{\text{DV}}>10$ GeV.
Specifically, the results for $m_{\phi}=30, 40,$ and 55 GeV are similar, and can already exclude Br$(h\to \phi\phi)$ as low as about $3\times 10^{-3}$ in the $c\tau_\phi \sim$ 1-10 mm range, with the current LHC Run-2 data.
In the HL-LHC phase, the reach can be enhanced by a corresponding factor of $3000/139 \sim 22$, touching $10^{-4}$, if the same level of background events can be achieved.
We note that in the upper plot of Fig.~\ref{fig:Br_ctau_lightscalar}, the curves for $m_\phi=30, 40,$ and 55 GeV are cut at the smallest $c\tau_\phi$ values as a result of a lack of statistics in this prompt-like regime.
Similar behavior will also arise in Fig.~\ref{fig:Br_mass_lightscalar}, for similar reasons.

The existing recent LHC searches targeting exactly this theoretical scenario in the same mass range include Refs.~\cite{CMS:2020iwv,CMS:2021yhb,ATLAS:2024qoo}.
Focusing on signatures of jets, these searches obtained bounds on Br$(h\to\phi\phi\to 4q)$, where $q$ labels quarks and anti-quarks.
We divide these bounds by the decay branching ratios of $\phi$ into a pair of the corresponding quarks obtained with HDECAY 3.4 in order to derive the bounds on Br$(h\to \phi\phi)$, and we find only the ATLAS search~\cite{ATLAS:2024qoo} with 140 fb$^{-1}$ integrated luminosity places competitive bounds compared to the recast ATLAS DV+jets search, while the other two searches' results are much weaker.
These bounds from Ref.~\cite{ATLAS:2024qoo} are shown as dot-dashed curves in Fig.~\ref{fig:Br_ctau_lightscalar} for $m_\phi=16, 40$, and 55 GeV.
We observe that their bounds for $c\tau_\phi\gtrsim 10$ mm are quite close to what we obtain with the Trackless-jet SR in the ATLAS DV+jets search.
However, for lower $c\tau_\phi$ values, our results are apparently much stronger.

Other LHC searches for such long-lived light scalars from the SM-like Higgs-boson decays include Refs.~\cite{CMS:2021juv,ATLAS:2022zhj,ATLAS:2019qrr,ATLAS:2019jcm}, but they are sensitive to even larger $c\tau_\phi$ values and are hence not considered here for the comparison purpose.

Since the ATLAS DV+jets search shows excellent sensitivities to relatively small $c\tau_\phi$ values, we should also consider comparison with existing bounds from prompt searches.
For instance, early LHC 13-TeV searches are summarized in Ref.~\cite{Cepeda:2021rql} that placed bounds on Br$(h\to \phi\phi\to XXYY)$ where $X$ and $Y$ labels final states include the SM $b$-quark, $\tau$ lepton, gluon, photon, and muon.
We find they correspond to bounds on Br$(h\to\phi\phi)$ at the best at the levels of 10\% - 100\%.
Similarly, the CMS search reported in Ref.~\cite{CMS:2024zfv} looks for the SM-like Higgs-boson decays to a pair of pseudoscalar bosons which all promptly decay to a pair of $b$-quarks, where the Higgs boson is supposed to be produced in association with a $Z$- or $W$-boson.
These results again only constrain Br$(h\to\phi\phi)$ in the order of 10\%, apparently weaker than those shown in Fig.~\ref{fig:Br_ctau_lightscalar}.
Once we include an efficiency factor requiring that the long-lived $\phi$ as considered here should decay promptly, these bounds should be even further weakened.
Therefore, we choose not to work out a recast of these prompt-search results, but only comment on their weak bounds as discussed here.

Finally, we choose not to show the current bounds on the invisible Higgs-boson decay branching ratio at 10.7\%~\cite{ATLAS:2023tkt}, since they would only be relevant for relatively large $c\tau_\phi$ values where the ATLAS DV+jets search is not or less sensitive.

In Fig.~\ref{fig:Br_mass_lightscalar} we choose a list of benchmark values of $c\tau_\phi=0.20, 0.40, 1$, and 10 mm, and present the numerical results in the $(m_\phi, \text{Br}(h\to\phi\phi))$ plane.
In general, we see that the sensitivity is enhanced quickly for increasing $m_\phi$ from $\sim 10$ GeV to 30 GeV, and then it roughly stabilizes for heavier $\phi$.
This can be understood easily from the selection cut of $m_{\text{DV}}>10$ GeV implemented in the ATLAS DV+jets search.

We comment that while we restrict ourselves to the $\phi$ decays to $b\bar{b}$, we could consider the decays into a pair of e.g.~down quarks only instead.
We find that the cut efficiencies with the $b\bar{b}b\bar{b}$ final state are similar to those with the $d\bar{d}d\bar{d}$ final state.
However, since in the $\mathcal{O}(10)$ GeV range of $m_{\phi}$, Br$(\phi\to d\bar{d})$ is orders of magnitude smaller than Br$(\phi \to b\bar{b})$, resulting in weaker bounds on Br$(h\to \phi\phi)$ in the end, we do not study this channel.

For completeness, we show in Appendix~\ref{appendix:bbbar} a plot for Br$(\phi\to b\bar{b})$ as a function of $m_\phi$, as obtained in HDECAY 3.4.

\subsection{Light scalars from a heavy Higgs boson}\label{subsec:results_heavyHiggs_lightscalar}

\begin{figure}[t]
    \centering
    \includegraphics[width=0.495\textwidth]{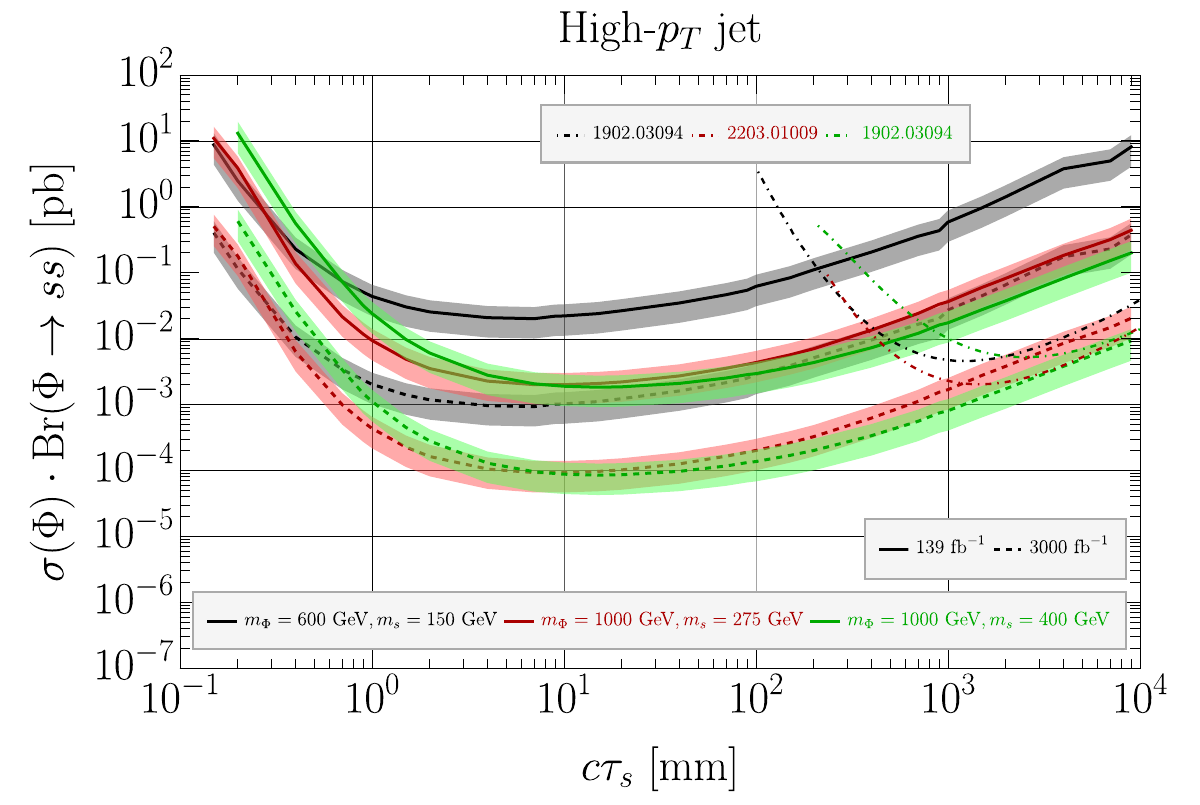}
    \includegraphics[width=0.495\textwidth]{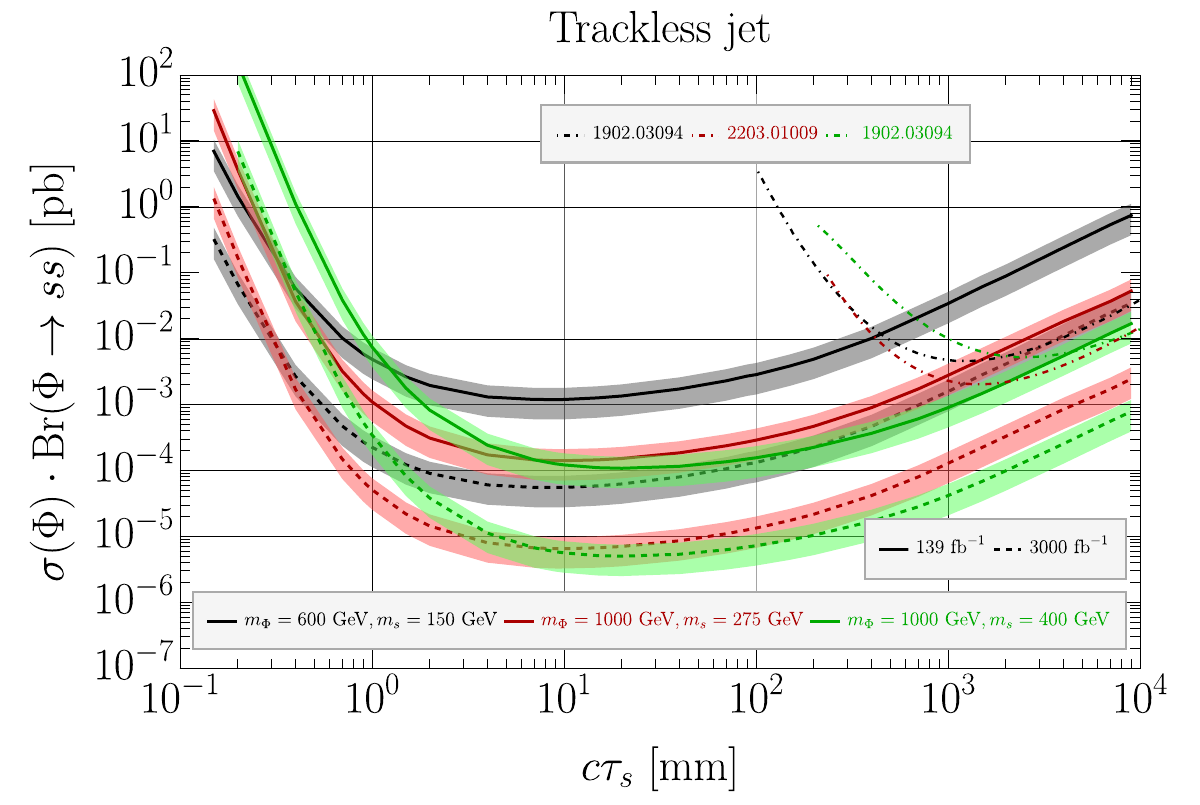}
    \caption{Sensitivity reach of the ATLAS DV+jets search to the long-lived light scalars $s$ produced from a non-SM heavy neutral boson $\Phi$, shown in the plane $\sigma(\Phi)\cdot\text{Br}(\Phi\to s s )$ vs.~$c\tau_s$, with the High-$p_T$-jet SR (upper plot) and the Trackless-jet SR (lower plot).  The three considered benchmark points are listed in the legend in the bottom of the plots.
    In each plot, the dot-dashed curves are existing bounds extracted from Ref.~\cite{ATLAS:2019qrr} for the black and green curves and from Ref.~\cite{ATLAS:2022zhj} for the red curve, respectively, with the colors corresponding correctly to the benchmark combinations of masses.}
    \label{fig:scalar_heavyHiggs}
\end{figure}

The numerical results for light scalars produced from a hypothetical heavy BSM scalar boson are shown in Fig.~\ref{fig:scalar_heavyHiggs}.
Again, we display two plots for the two SRs of the ATLAS DV+jets search, respectively.
Both plots are displayed in the $\sigma(\Phi)\cdot\text{Br}(\Phi\to ss)$ vs.~$c\tau_s$ plane for the three benchmark points of the masses of $\Phi$ and $s$.
Solid and dashed curves are bounds corresponding to 139 fb$^{-1}$ and 3000 fb$^{-1}$ integrated luminosities.
We first observe that the Trackless-jet SR performs much better than the High-$p_T$-jet SR in general, except when the lifetime of the $s$ particle is so short that fewer events have displaced jets passing the jet-selection requirements of the Trackless-jet SR, cf.~Table \ref{tab:event_acceptance}.
Then, we find for the $(m_\Phi=600\text{ GeV}, m_s=150\text{ GeV})$ benchmark the ATLAS DV+jets search is less powerful than for the other two benchmarks with heavier $\Phi$ and $s$.
This is, as we explained above in Sec.~\ref{subsec:results_lightscalar}, primarily a result of the harder jets stemming from heavier Higgs bosons and light scalars.

Further, we compare these results with existing bounds for the same benchmarks obtained in some ATLAS searches for displaced hadronic jets~\cite{ATLAS:2019qrr,ATLAS:2022zhj}, which are plotted in the dot-dashed line style.
Ref.~\cite{ATLAS:2019qrr} provides the leading bounds on the benchmarks of $(m_\Phi=600\text{ GeV}, m_s=150\text{ GeV})$ and $(m_\Phi=1000\text{ GeV}, m_s=400\text{ GeV})$, with 33.0 fb$^{-1}$ integrated luminosity of data collected during the LHC Run 2, while Ref.~\cite{ATLAS:2022zhj} shows the constraints on the benchmark $(m_\Phi=1000\text{ GeV}, m_s=275\text{ GeV})$ with the full LHC Run-2 data of 139 fb$^{-1}$.
We arrive at similar conclusions to those reached in Sec.~\ref{subsec:results_lightscalar} that the considered ATLAS DV+jets search can be particularly sensitive to $c\tau_s$ ranging roughly between 0.1 and 100 mm. 
For these scenarios, no existing prompt-search bounds exist to our knowledge, and therefore a corresponding comparison is not made here.

We note that bounds for the benchmarks $(m_\Phi=600\text{ GeV}, m_s=150\text{ GeV})$ and $(m_\Phi=1000\text{ GeV}, m_s=400\text{ GeV})$ were also attained in another ATLAS search for displaced hadronic jets reported in 2019~\cite{ATLAS:2019jcm}, but these results are too weak and hence not shown here.
Also, the limits for the benchmark $(m_\Phi=600\text{ GeV}, m_s=150\text{ GeV})$ obtained in Ref.~\cite{ATLAS:2022zhj} are actually slightly stronger than those from Ref.~\cite{ATLAS:2019qrr} which we choose to show here, but they are valid only for a narrower $c\tau_s$ range, and therefore also not presented in Fig.~\ref{fig:scalar_heavyHiggs}.

\subsection{Heavy neutral leptons from SM-like Higgs boson decays}~\label{subsec:results_HNL_SMHiggs}

\begin{figure}[t]
    \centering
    \includegraphics[width=0.495\textwidth]{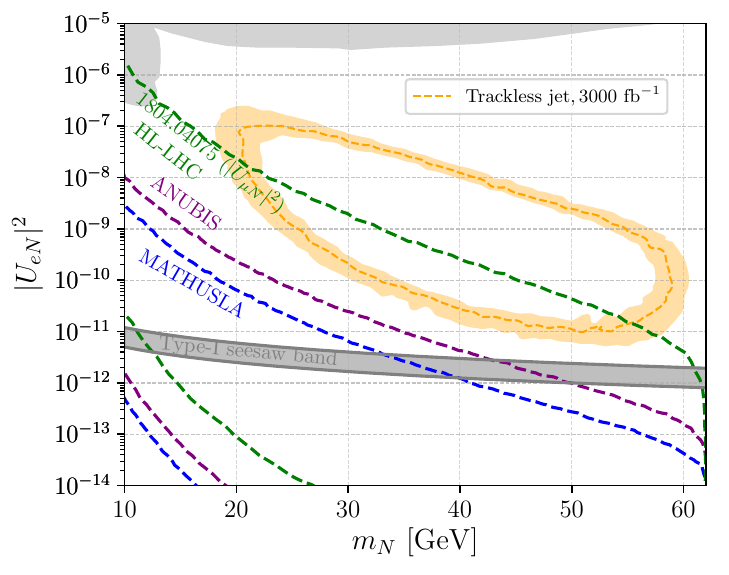}
    \caption{Sensitivity reach at 95\% C.L.~of the ATLAS DV+jets search to the $U(1)_{B-L}$ scenario with the Trackless-jet SR and 3000 fb$^{-1}$ integrated luminosity, assuming the same level of background events as that with 139 fb$^{-1}$.
    The error band accounts for an uncertainty of 50\% and the region inside the orange closed curve is the one that the Trackless-jet SR of the search can exclude at 95\% C.L.~by the HL-LHC projection.
    The MATHUSLA and ANUBIS sensitivity reach for 3000 fb$^{-1}$ integrated luminosity is retrieved from Ref.~\cite{Hirsch:2020klk}.
    The dark-gray band labeled with ``Type-I seesaw band'' corresponds to the region which can be explained by the Type-I seesaw relation $|U_{eN}|^2  \simeq  m_{\nu_e}/m_N$, for $m_{\nu_e}$ between 0.05 and 0.12 eV.
    The upper light-gray region is the currently excluded parameter space in the minimal HNL scenario where the mixing parameter mediates both the production and decay of the Majorana HNL, obtained in Refs.~\cite{ATLAS:2022atq,CMS:2022fut,CMS:2024xdq}.
    The dashed green curve is the expected sensitivity reach at the HL-LHC with a search strategy of a single DV that includes a muon, and is valid for an HNL mixed with the SM muon neutrino only~\cite{Deppisch:2018eth}; nevertheless, it is shown here for comparison purpose.
    }
    \label{fig:HNL_U1BmL}
\end{figure}

For the scenarios of the HNLs decaying semi-leptonically in the $U(1)_{B-L}$ model, we find that for the two choices of the integrated luminosity and the two available SRs, only the Trackless-jet SR with 3000 fb$^{-1}$ integrated luminosity can be sensitive.
We present the numerical results in the plane $|U_{eN}|^2$ vs.~$m_N$, in Fig.~\ref{fig:HNL_U1BmL}.
The ATLAS DV+jets search can probe the parameter regions roughly between 20 and 60 GeV for $m_N$ and between $10^{-11}$ and $10^{-7}$ for $|U_{eN}|^2$, for the total data collected during the HL-LHC.
This is in contrast with the predicted sensitivity reach of the proposed LHC far detectors ANUBIS~\cite{Bauer:2019vqk} and MATHUSLA~\cite{Chou:2016lxi,Curtin:2018mvb,MATHUSLA:2020uve}, derived in Ref.~\cite{Hirsch:2020klk} for an integrated luminosity of 3000 fb$^{-1}$ and the same production channel of the HNLs.
Such far detectors are sensitive to complementary parts of the parameter space, as shown in Fig.~\ref{fig:HNL_U1BmL}, that correspond to longer lifetimes with a lower mass and a smaller mixing parameter.
Similarly, an estimate performed in Ref.~\cite{Deppisch:2018eth} for the HL-LHC sensitivity reach to $|U_{\mu N}|^2$ as a function of $m_N$ is also overlaid here (despite the different mixing parameter considered), as a dashed green curve.
Ref.~\cite{Deppisch:2018eth} proposed a search strategy of a single DV that includes a muon and its results are hence valid only for an HNL that mixes with the SM muon neutrino $\nu_\mu$.
The shown results were obtained under the optimistic assumption of vanishing background after the selection requirements are imposed.
We comment that if we consider an HNL that mixes with $\nu_\mu$ instead of $\nu_e$, the ATLAS DV+jets search recast would give similar projected sensitivity reaches in the relevant mass range, and therefore, we depict these results from Ref.~\cite{Deppisch:2018eth} here, in order to show that the ATLAS DV+jets search can probe relatively prompt-like regions in the parameter space compared to other LHC DV searches.
We, in addition, display a dark-gray band in the plot for the parameter region that can explain the active-neutrino mass with the Type-I seesaw relation $|U_{eN}|^2  \simeq  m_{\nu_e}/m_N$ for $m_{\nu_e}$ between 0.05 eV and 0.12 eV corresponding to neutrino-oscillation~\cite{Canetti:2010aw} and cosmology observations~\cite{Planck:2018vyg}, respectively.
Finally, we show in the light-gray parts of the plot the current bounds from LHC searches~\cite{ATLAS:2022atq,CMS:2022fut,CMS:2024xdq} on $|U_{eN}|^2$ for Majorana HNLs in the minimal scenario, where both the production and decay of the HNLs are induced only by the mixing.

Before closing, we briefly explain the boundary of the sensitive region.
The lower and upper mass reach is determined, respectively, by the $m_{\text{DV}}>10$ GeV cut and the kinematic threshold $m_h/2$.
For parameter regions with heavy $N$ and large mixing $|U_{eN}|^2$, the HNL is promptly decaying before reaching $R_{xy}=4$ mm, while for light $N$ with small $|U_{eN}|^2$ the HNL is so long-lived that it decays only after leaving $R_{xy}=300$ mm and $|z|=300$ mm (see Sec.~\ref{sec:search}).

\section{Conclusions}\label{sec:conclusions}

ATLAS has published an LLP search for displaced vertices and jets~\cite{ATLAS:2023oti}.
It has been recast and reinterpreted in long-lived axion-like particles from top-quark decays in Ref.~\cite{Cheung:2024qve} .
In this work, we have applied this recast in order to reinterpret the bounds in terms of various possible LLP candidates that are produced from decays of the SM (or a BSM) Higgs boson and decay hadronically.

Specifically, we have studied the following three theoretical scenarios: 1) long-lived light scalars $\phi$ produced from the SM-like Higgs-boson $h$ decays and decaying into a pair of $b$-quarks, 2) long-lived light scalars $s$ originating from a heavy BSM Higgs boson $\Phi$'s decays and decaying into $b\bar{b}$, $c\bar{c}$, or $\tau\bar{\tau}$, and 3) long-lived HNLs $N$ from the decays of the SM-like Higgs boson in the framework of a $U(1)_{B-L}$ model.
Besides the exclusion bounds corresponding to the integrated luminosity of 139 fb$^{-1}$ made use of in the ATLAS DV+jets search, we also project future HL-LHC sensitivity reach at 95\% C.L.~with 3 ab$^{-1}$ data, assuming the same number of background events.
In the first scenario, we have presented the numerical results in the $(c\tau_\phi, \text{Br}(h\to \phi\phi))$ and $(m_\phi, \text{Br}(h\to \phi\phi))$ planes.
We find the Trackless-jet SR shows more promising sensitivities than the High-$p_T$-jet SR.
This is mainly due to the relatively small mass of the SM-like Higgs boson, compared to the jet-selection requirements of the ATLAS DV+jets search, and to the stronger jet-$p_T$ requirements of the High-$p_T$-jet SR.
Compared to other published LLP or prompt searches at the LHC, we find the ATLAS DV+jets search can extend the reach to Br$(h\to \phi\phi)$ largely in the $\mathcal{O}(1)$ mm range.
In the second theoretical scenario, we show results in the $\sigma(\Phi)\cdot\text{Br}(\Phi \to ss)$ vs.~$c\tau_s$ plane, for a fixed set of benchmarks of the masses of $\Phi$ and $s$ following other published ATLAS searches.
Owing to the larger masses of $\Phi$ and $s$ compared to those of $h$ and $\phi$ in the previous scenario, we find stronger analysis cutflow efficiencies.
Both the High-$p_T$-jet and Trackless-jet SRs with 139 fb$^{-1}$ integrated luminosity can exclude values of $\sigma(\Phi)\cdot\text{Br}(\Phi \to ss)$ orders of magnitude smaller than the current bounds for the relevant $c\tau_s$ range, and can probe $\sigma(\Phi)\cdot\text{Br}(\Phi \to ss)$ for even smaller $c\tau_s$ values, especially in the range of 1-100 mm.
The final theoretical scenario concerns a UV-complete $U(1)_{B-L}$ model and we show the numerical results in the planes spanned by model parameters $m_N$ and $|U_{eN}|^2$.
Here, with the current bounds on the $U(1)_{B-L}$ model parameters, only the HL-LHC with 3000 fb$^{-1}$ integrated luminosity can probe unexcluded parameter space with the Trackless-jet SR in the ATLAS DV+jets search, covering regions of the parameter space complementary to those to which other DV searches at the HL-LHC and future LHC far detectors such as MATHUSLA and ANUBIS could be sensitive.

In summary, we find that the ATLAS DV+jets search is particularly powerful at testing LLPs with a proper decay length of about 1-100 mm, which is a unique region that is hard to access by either the prompt searches or other LLP searches at the LHC.
With this paper, we thus motivate further works probing LLPs of a relatively small lifetime with the ATLAS DV+jets search.

\section*{Acknowledgements} 
We thank Kingman Cheung, Giovanna Cottin, Martin Hirsch, and Wei Liu for useful discussions.

\vspace{1cm}

\appendix
\section{$\phi$ decay branching ratio into $b\bar{b}$ as a function of $m_{\phi}$}\label{appendix:bbbar}

\begin{figure}[H]
    \centering
    \includegraphics[width=0.495\textwidth]{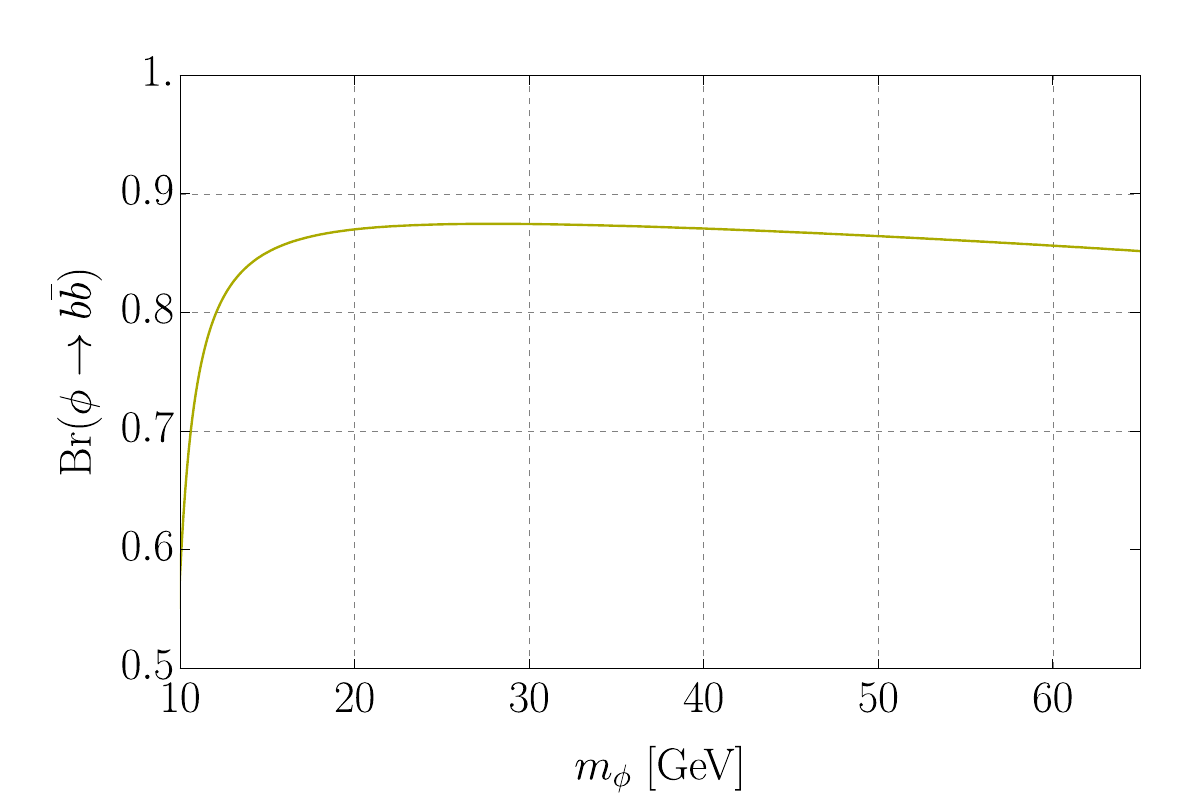}
    \caption{Decay branching ratio of the light scalar particle $\phi$ into a pair of $b$-quarks, as a function of its mass $m_\phi$.}
    \label{fig:phi2bbbar_br}
\end{figure}

Here, for reference, we show the decay branching ratio of $\phi$ into a pair of $b$-quarks as a function of $m_\phi$ in Fig.~\ref{fig:phi2bbbar_br}, extracted from the computation results given by HDECAY 3.4~\cite{Djouadi:1997yw,Djouadi:2018xqq}.
This concerns the numerical results of the first theoretical scenario.

\bibliography{refs}

\end{document}